\documentclass[preprintnumbers,article,amsmath,amssymb,floatfix,10pt,prd,onecolumn,
superscriptaddress,nofootinbib]{revtex4-2}
\usepackage{bbm}
\usepackage{amsfonts}
\newcommand{\levicivita}{}
\def\levicivita#1#{\tensor#1{\epsilon}}

\usepackage{mathrsfs}
\usepackage{latexsym}
\usepackage[colorlinks=true, citecolor=blue, urlcolor = blue, linkcolor= red, bookmarks=true]{hyperref}
\usepackage{epsfig}
\usepackage{epstopdf}
\usepackage{graphicx}
\usepackage{amssymb}
\usepackage{dcolumn}
\usepackage{bm}
\usepackage{color}
\usepackage{comment}
\usepackage{xcolor}
\usepackage{amstext}
\usepackage{centernot}

\usepackage{subfigure}
\usepackage{mathrsfs}
\usepackage{eqnarray}
\usepackage{amsmath}

\usepackage{xspace}

\usepackage[export]{adjustbox}
\usepackage{float}
\usepackage{csquotes}
\usepackage[colorinlistoftodos]{todonotes}
\usepackage{hyperref}
\begin{document}
\title{Reconstruction of symmteric teleparallel gravity with energy conditions }
\author{Irfan Mahmood}
\email{mahirfan@yahoo.com}\affiliation{Department of Mathematics, Shanghai University  and Newtouch Center for Mathematics of Shanghai University,  Shanghai ,200444, P.R.China.}\affiliation{Centre for High Energy Physics, University of the Punjab, 54590, Lahore, Pakistan.}
\author{Hira Sohail}
\email{hirasohail02@outlook.com}\affiliation{Centre for High Energy Physics, University of the Punjab, 54590, Lahore, Pakistan.}
\author{Allah Ditta}
\email{mradshahid01@gmail.com}\affiliation{Department of Mathematics, Shanghai University  and Newtouch Center for Mathematics of Shanghai University,  Shanghai ,200444, P.R.China.}
\author{S. H. Shekh}
\email{da\_salim@rediff.com}
\affiliation{Department of Mathematics. S. P. M. Science and Gilani Arts Commerce
College, Ghatanji, Dist. Yavatmal,\\ Maharashtra-445301, India.}
\author{Anil Kumar Yadav}
\email{abanilyadav@yahoo.co.in} \affiliation{Department of Physics, United College of Engineering and Research, Greater Noida – 201 306, India}
\begin{abstract}
\textbf{Abstract:} This research investigates the impact of modified gravity on cosmic scales, focusing on $f(Q)$ cosmology. By applying energy conditions, the study reconstructs various $f(Q)$ models, considering an accelerating Universe, quintessence, and a cosmological constant $\Lambda$. Using up-to-date observational data, including the Supernova Pantheon sample and cosmic chronometer data, Hubble constants $H_0$ are estimated as $70.37^{+0.84}_{-0.92}$ km/sec/Mpc (from $H(z)$ data) and $70.02^{+0.44}_{-0.25}$ km/sec/Mpc (from pantheon compilation of SN Ia data). The matter energy density parameter ($\Omega_{0m}$) is calculated as $0.26^{0.015}_{-0.010}$(OHD) and $0.27^{0.025}_{-0.014}$(SN Ia). Furthermore, as a function of redshift $z$, explicit expressions of $f(Q)$ and the EOS parameter $\omega$ are produced, and their graphical analysis describes the late time acceleration of the Universe without the usage of dark energy.
\end{abstract}

\maketitle

\date{\today}

\section{Introduction}
The scientific purpose of ongoing and future cosmological surveys is to comprehend the true nature of cosmic acceleration, which is based on evaluating the conventional cosmological model $\lambda$, cold-dark-matter ($\Lambda$CDM), and any deviation from it. Keeping the cosmological constant as the primary source of this phenomenon in mind, one can build gravity theories that are indistinguishable from $\Lambda$CDM at the background level yet exhibit fascinating and identifiable signatures on the dynamics of perturbations. Following that, we will look into whether there is a gravity theory with these characteristics that can challenge the $\Lambda$CDM scenario.Cosmologists have recently become interested in modified gravity theories in order to better comprehend the origins and purpose of dark energy. The origin of dark energy is regarded as a modification of gravity in modified gravity. Much study has revealed that modified gravity theories can explain both early and late time acceleration of the cosmos. As a result, there are numerous motives to uncover theories other than the traditional formulation of GR. Several modified theories, such as $f(R)$ theory \cite{B1,B2}, $f (T)$ theory \cite{B3,B4,B5}, $f (T, B)$ theory \cite{B6}, $f (R, T)$ theory \cite{B7,B8}, $f (Q, T)$ theory \cite{B9}, $f (G)$ theory \cite{B10} and $f (R, G)$ theory \cite{B11,B12} have been proposed in the literature.
\\

We will consider an extension of the Symmetric Teleparallel General Relativity, the f(Q)-gravity.f(Q) gravity \cite{A1,A2,A3,A4,A5} is a recent MG theory that has received a lot of attention. It is a member of the Symmetric Teleparallel Gravity \cite{A6,A7,A8} family, in which gravity is attributed to non-metricity and $f(Q)$ is a generic function of the non-metricity scalar, $Q$.
The study of $f(Q)$ gravity has advanced rapidly, leading to fascinating applications \cite{C1,C2,C3,C4,C5,C6,C7,C8,C9,C10,C11,C12,C13,C14,C15}. When we look at the history of the universe's expansion, we can see that several cosmological parameters play a significant part in defining the cosmological model's cosmic evolution. The energy conditions, as is well known, represent methods for implementing the positiveness of the stress-energy tensor in the presence of matter. They can also be used to characterize the attractive nature of gravity, in addition to attributing the underlying causal and geodesic structure of space-time \cite{B6}.
\\

Various reconstruction techniques have been utilized in recent years to explore some cosmologically interest elements of the late cosmos.The full analysis of the reconstruction program for general scalar-Gauss-Bonnet gravity utilizing high-order string corrections is given  in \cite{BB1}. Similarly, using the reconstruction techniques in \cite{BB2}, many functions that satisfy cosmological models are reconstructed, as well as the equations giving energy conditions for the reconstructed $f(\mathcal{G},\phi,(\Delta \phi)^2,T)$ gravity. Recently, in \cite{BB3}, the reconstruction technique is utilized to obtain explicit models of the $f(Q,T)$ Lagrangian for various types of matter sources and Einstein's static universe. Using the same approach, we investigated the strong, weak, null, and dominant energy conditions for reconstructing $f(Q)$ gravity models using a matter content to be a holographic dark energy source  in this research. Our Universe's true acceleration phase, limitations that the strong energy requirement should be breached. This limitation, along with the real Hubble and deceleration parameters, allowed us to explore the viability of various kinds of f (Q) gravity.\\

One fundamental question in modified gravities is determining the unknown function that enters the theory.Although some general aspects, such as instabilities or the existence of Noether symmetries, can be determined theoretically, the most potent technique is the utilization of observational data \cite{HH1,HH2,HH3,HH4,HH5,HH6}.
As a result, we are interested in utilizing expansion data in this work, such as supernovae type Ia data (SNIa) and Hubble cosmic datasets.
The following is how the current paper is organized: In Section \ref{1}, we give the $f(Q)$ gravity formalism and accompanying modified Friedmann equations. In Sec.\ref{3}, we constrained the model parameters in Sec. \ref{4} by utilizing 55 points from the Hubble dataset and 1048 points from the SNe Ia dataset. In Section \ref{4}, we recreated Teleparallel gravity using various enegry conditions and estimated the EoS parameter for each of them.Finally, in Section \ref{5}, we discuss the results with concluding remarks.

\section{Basics of $f(Q)$ gravity wity field equations}\label{1}
Let us start with the following action of $f(Q)$ gravity given by \cite{SH1}
\begin{equation}\label{eq.1}
    S=\Bigg[\frac{1}{2}\int f(Q)+\int L_m \Bigg] \sqrt{-g} d^4 x, 
\end{equation}
where $f(Q)$ is a generic function of $Q$, $L_m$ is the matter Lagrangian density, and $g$ is the metric $g_{\mu \nu}$ determinant.
\begin{equation}\label{eq.2}
    Q_{\gamma \mu \nu}=\Delta_\gamma g_{\mu \nu},
\end{equation}
\begin{equation}\label{eq.3}
    Q_\gamma=Q^{\hspace{0.2cm}\mu}_{\gamma \hspace{0.3cm} \mu} \hspace{1cm}\Tilde{Q}_\gamma=Q^{\mu}_{\hspace{0.2cm}\gamma \mu},
\end{equation}
Furthermore, as a function of the nonmetricity tensor, the superpotential is given by
\begin{equation}\label{eq.4}
    4P^{\gamma}_{\hspace{0.2cm}\mu \nu}=-Q^{\gamma}_{\hspace{0.2cm}\mu \nu}+2Q_{(\mu^{\gamma}\nu)}-Q^{\gamma}g_{\mu \nu}-\Tilde{Q}^{\gamma}g_{\mu \nu}-\delta^{\gamma} _{\hspace{0.2cm}(\gamma}Q_{\nu )},
\end{equation}
where the trace of nonmetricity tensor (\ref{eq.4}) has the form

\begin{equation}\label{eq.5}
    Q=-Q_{\gamma \mu \nu}P^{\gamma \mu \nu},
\end{equation}

Another important component of our approach is the matter's energy- momentum tensor, which is defined as

\begin{equation}\label{eq.6}
    T_{\mu \nu}=-\frac{2}{\sqrt{-g}}\frac{\delta(\sqrt{-g}L_m)}{\delta g^{\mu \nu}},
\end{equation}
Using the variation of action (\ref{eq.1}) with respect to the metric tensor, one can get the field equations as
\begin{equation}\label{eq.7}
    \frac{2}{\sqrt{-g}}\Delta_\gamma \bigg(\sqrt{-g} f_Q P^{\gamma}_{\hspace{0.2cm}\mu \nu}\bigg)+\frac{1}{2}g_{\mu \nu}f+f_Q \bigg(P_{\mu \gamma i} Q^{\hspace{0.15cm}\gamma i }_\nu -2 Q_{\gamma i \mu}P^{\gamma i }_{\hspace{0.4cm}\nu}\bigg)=-T_{\mu \nu},
\end{equation}
where $f_Q = \frac{df}{dQ}$. Furthermore, we can take the variation of (\ref{eq.1}) with regard to the connection, resulting in
\begin{equation}\label{eq.8}
    \Delta_\mu \Delta_\gamma \Bigg(\sqrt{-g}f_Q P^{\gamma}_{\hspace{0.2cm}\mu \nu}\Bigg)=0,
\end{equation}
Let us Consider the Friedman-Lemâitre-Robertson-Walker (FLRW) line-element of the form, which is spatially homogeneous and isotropic.
\begin{equation}\label{eq.12}
ds^2=dt^2-a^2(t)\Big[dr^2+r^2(d\theta^2+\sin^2{\theta}\hspace{0.1cm} d\phi^2)\Big],
\end{equation}
where $a$ is scale factor that defines expansion rate of the Universe and $(t, r,\theta,\phi)$ are the co-moving coordinates.\\

For the FLRW space - time, we obtain the non-metricity scalar as $Q = 6H^{2}$. We consider the matter content of the derived Universe model as consisting of standard perfect fluid matter whose energy momentum tensor is given by 
\begin{equation}\label{emt}
T_{\mu\nu} = (\rho + p)u_{\mu}u_{\nu} + pg_{ij}
\end{equation}
where $\rho$ and $p$ represent energy density and pressure of the perfect fluid respectively. $u_{\mu}$ is the four velocity vector satisfying $u^{\mu}u_{\mu} = -1$.\\

The modified Friedmann equations for $f (Q)$ gravity can be found by solving the equation of motion (\ref{eq.7}) for the FLRW line-element (\ref{eq.12}) with the fluid of stress-energy tensor (\ref{emt}).

\begin{equation}\label{eq.13}
    3H^2=\frac{1}{2f_Q}\Bigg(-\rho+\frac{f}{2}\Bigg),
\end{equation}

\begin{equation}\label{eq.14}
    \Dot{H}+3H^2+\frac{\Dot{f_Q}}{f_Q}H=\frac{1}{2f_Q}\Bigg(p+\frac{f}{2}\Bigg),
\end{equation}
The differentiation with regard to cosmic time $t$ is represented by the dot over the parameters.
Furthermore, the modified Friedmann equations allow us to write the density and pressure of the Universe as
\begin{equation}\label{eq.15}
    \rho=\frac{f}{2}-6 H^2f_Q,
\end{equation}

\begin{equation}\label{eq.16}
p=\Bigg(\Dot{H}+3H^2+\frac{\Dot{f_Q}}{f_Q}H\Bigg)(2f_Q)-\frac{f}{2},    
\end{equation}
The standard matter distribution in $f(Q)$ theory of gravity validates the energy conservation equation \cite{Mandal/2023}
\begin{equation}\label{ec-1}
 \frac{d\rho}{dt} + 3H(1+\omega)\rho = 0   
\end{equation}
In Eq. (\ref{ec-1}), the Equation of state (EoS) parameter $\omega$ that characterizes the relationship between the pressure and energy density of a fluid in the Universe. It plays a crucial role in understanding the dynamics of the expansion of the Universe and the nature of the components of energy content of the Universe. It is defined as
\begin{equation}\label{eq.21}
\omega=\frac{p}{\rho}
\end{equation}
The Hubble parameter and deceleration parameters are read as
\begin{equation}\label{eq.17}
H=\frac{\Dot{a}}{a}, \hspace{1cm}q=-\frac{1}{H^2}\frac{\Ddot{a}}{a},
\end{equation}
These parameters allow us to represent time derivatives of $H$ as,
\begin{equation}\label{eq.18}
\Dot{H}=-H^2(1+q)
\end{equation}
Using Eq. (\ref{eq.18}), Eqs. (\ref{eq.15}) and (\ref{eq.16}) can be rewritten as
\begin{equation}\label{eq.19}
\rho=\frac{f}{2}-6H^2f_Q
\end{equation}
\begin{equation}\label{eq.20}
p = 2f_Q\Bigg(-H^2(1+q)+3H^2+\frac{\Dot{f_Q}}{f_Q}H\Bigg)
\end{equation}
Putting the values of $\rho$ and $p$ from Eqs.(\ref{eq.15}) and (\ref{eq.16}) in Eq. (\ref{eq.21}), we obtain, 
\begin{equation}\label{eq.w}
  \omega=-1-\frac{2 f_Q \Dot{H}+2 H \Dot{f_Q}}{6 H^2 f_Q-f/2}
\end{equation}
Since, we have considered perfect fluid as the matter content of the Universe in the framework of symmetric teleparallel gravity therefore, to determine expansion rate $(1+z)H(z) = -\frac{dz}{dt}$, we define the following functional from of $E(z)$ 
\begin{equation}\label{H}
 E(z) = \frac{H^{2}(z)}{H_{0}^{2}} = \Omega_{0m}(1+z)^3+ \Omega_{0Q}   
\end{equation}
where $H(z)$, $H_{0}$ and $\Omega_{0m}$ denote the Hubble parameter, present value of Hubble constant and present value of matter energy density parameter respectively. It is worthwhile to note the $\Omega_{0Q}$ denotes the energy density parameter that generate due to geometry of $f(Q)$ gravity. Moreover, in Sahni et al \cite{Sahni/2003}, the functional form of $E(z)$ is read as
\begin{equation}\label{E}
 E(z) = \Omega_{0m}(1+z)^{3} + \alpha(1+z)^{2} + \beta(1+z) + \mu   
\end{equation}
It is worthwhile to note that at $z = 0$, $H(z) = H_{0}$, therefore the validation of Eq. (\ref{E}) restricts that $E(z) = 1$, at $z = 0$, which constraints the relationship between constants $\alpha$, $\beta$ and $\mu$ as $\alpha + \beta + \mu = 1 - \Omega_{0m}$. To fulfill this requirement, one may consider the simplest functional form of $E(z)$ as given in Eq. (\ref{H}) with restriction $\Omega_{0Q} = 1 - \Omega_{0m}$. Thus, we rewrite Eq. (\ref{H}) as,
\begin{equation}\label{eq.H}
 H(z)  = H_{0}\left[\Omega_{0m}(1+z)^3+(1-\Omega_{0m})\right]^{\frac{1}{2}}  
\end{equation}
In Lohakare et al. \cite{Lohakare/2022}, the authors have analyzed the geometrical and dynamical parameters of modified teleparallel - Gauss - Bonnet model and constrained $\Omega_{0m}$ and other constants by fitting the model with observation Hubble data and Pantheon compilation of SN Ia data. Lohakare et al. \cite{Lohakare/2022} have considered $H _{0} = 70.7$ km/s/Mpc. In this research, we have constrained both $H_{0}$ and $\Omega_{0m}$ by fitting the experimental data and their measurements. Furthermore, we also regenerate a symmetric teleparallel gravity theory by employing energy conditions. 
\section{Observational Constraints}\label{3}
In this section, we describe observational $H(z)$ data (OHD) and Pantheon compilation of Supernovae Type Ia (SN Ia) data. We also described the statistical methodology for constraining model parameters $H_{0}$ and $\Omega_{m0}$ .
\begin{itemize}
\item {\bf OHD}: We have taken over $55~H(z)$ observational data points in the range of $0\leq z\leq 2.36$, obtained from cosmic chronometric technique. These all $55~H(z)$ data points are compiled in table II of Ref. \cite{Lohakare/2022}.\\

\item {\bf SN Ia}: We have used Pantheon compilation of SN Ia data \cite{Scolnic/2018} which includes 1048 SN Ia apparent magnitude measurements in the redshift range $0.01 < z < 2.3$. We also note that the Pantheon compilation of SN Ia data is comprising 40 binned data points in the redshift region $0.0014 \leq z \leq 1.62$ \cite{Scolnic/2018}. The redshift range of $0.01\leq z \leq 2.3$ is supported by 1048 apparent magnitude measurements $m_{B}$ in this data set. \\
\end{itemize}
The $\chi^{2}$ is read as 
\begin{equation}
\label{chi1}
\chi^{2} = \sum_{i=1}^{N}\left[\frac{E_{th}(z_{i})- E_{obs}(z_{i})}{\sigma_{i}}\right]^{2}
\end{equation}
where $E_{th}(z_{i})$ and $E_{obs}(z_{i})$ denote the theoretical values and observed values of corresponding parameters respectively. $\sigma_{i}$ and N are standard errors in $E_{obs}(z_{i})$ and number of data points.\\
\begin{figure}[ht]
\centering
\includegraphics[width=6.5cm,height=6.5cm,angle=0]{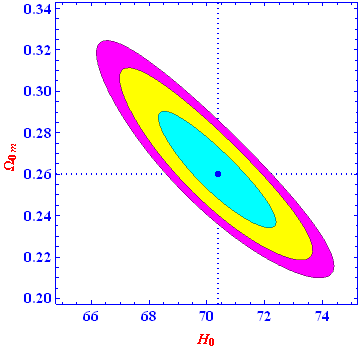}
\caption{Two dimensional contours at $1\sigma$, $2\sigma $ and $3\sigma$ confidence regions by bounding our model with latest 46 observational Hubble data. The unit of $H_{0}$ is $\;km\;s^{-1}\;Mpc^{-1}$}.
\end{figure}  
\begin{figure}[ht]
\centering
\includegraphics[width=6.5cm,height=6.5cm,angle=0]{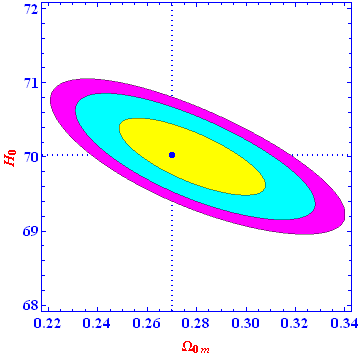}
\caption{Two dimensional contours at $1\sigma$, $2\sigma $ and $3\sigma$ confidence regions by bounding our model with Pantheon compilation of SN Ia data. The unit of $H_{0}$ is $\;km\;s^{-1}\;Mpc^{-1}$}.
\end{figure}  

Fig. 1 and Fig. 2 depict the two dimension contours at $1\sigma$, $2\sigma $ and  $3\sigma$ confidence levels of this model for observational Hubble data and Pantheon compilation of SN Ia data. We estimate Hubble constant as $H_{0} = 70.37^{+0.84}_{-0.92}$ km/sec/Mpc and $H_{0} = 70.02^{+0.44}_{-0.25}$ km/sec/Mpc with observational H(z) and SN Ia data respectively. Furthermore, we obtained the matter energy density parameter as $\Omega_{0m} = 0.26^{+0.015}_{-0.010}$ (OHD) and $\Omega_{0m} = 0.27^{+0.025}_{-0.014}$ (SN Ia). It is worthwhile to note that after knowing the value of $H_{0}$, one can easily estimate the present age of the Universe as following
$$H_0 (t_0-t) = \int_{0}^{z}\frac{dz}{(1+z) h(z)};~ h(z)=H(z)/H_{0}$$\\
where, $$H_0 t_0 =\lim_{z\rightarrow\infty}\int_{0}^{z}\frac{dz}{(1+z) h(z)}.$$
\begin{figure}[ht!]
\centering
\includegraphics[width=7cm,height=7cm,angle=0]{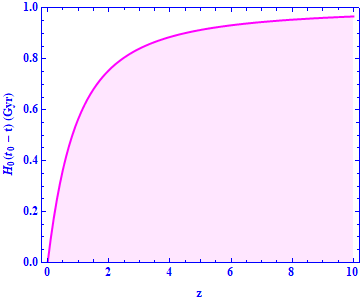}  
\caption{Plot of $H_{0}(t_{0} - t)$ versus redshift $z$.}
\end{figure}
Fig. 3 shows the variation of $H_{0}(t_{0} - t)$ versus of redshift $z$, where $t_{0}$ represents the age of the Universe at present epoch. For infinitely large value of $z$, we obtain $H_{0}t_{0} \sim 1.003$ which in turn implies that $t_{0} \sim 1.003 H_{0}^{-1}$. In the derived Universe model, we obtained the present age of the universe as $t_{0} = 14.53^{+20.0}_{-0.17}$ Gyrs (OHD) and $t_{0} = 14.61^{+0.05}_{-0.10}$ Gyrs (SN Ia). In our previous investigation, we obtained the present age of the Universe as $t_{0} = 14.04^{+0.33}_{-0.34}$ Gyrs \cite{Yadav/2021PRD}. Some estimations on the present age of the Universe, close to age of the Universe in this research, are also given in Refs. \cite{Yadav/2021PDU,Goswami/2021,Prasad/2020}. \\
\begin{figure}[ht!]
\centering
\includegraphics[width=7cm,height=7cm,angle=0]{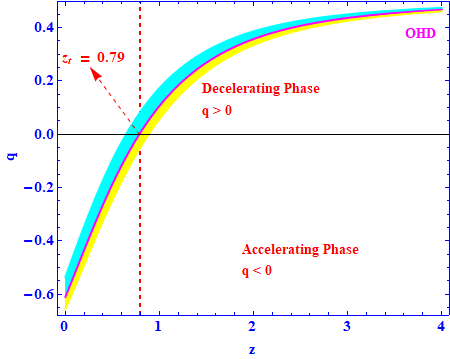}
\includegraphics[width=7cm,height=7cm,angle=0]{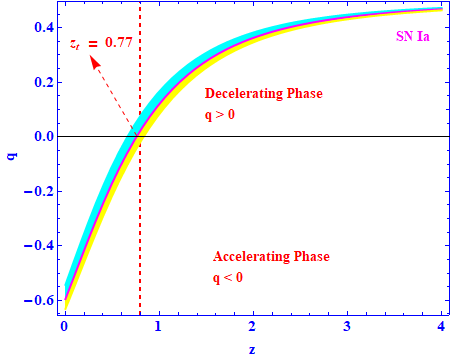}  
\caption{Plot of $q$ versus redshift $z$ for OHD (left panel) and Pantheon compilation of SN Ia data (right panel).}
\end{figure}
The deceleration parameter $q = -1 + \frac{(1+z)dH/dz}{H} =-1 + \frac{3 (1+z)^3 \Omega_{m0} }{2 \left[(1+z)^3 \Omega_{m0} -\Omega_{m0} +1\right]}$ describes the rate of expansion of the Universe. It is worthwhile to note that the positive $q$ resembles the decelerating phase of the the Universe while negative $q$ corresponds to the accelerating phase of the Universe. In this model, the dynamical behaviour of $q$ tells that the Universe was in decelerating mode of expansion with positive $q$ and it turn into accelerating mode with negative $q$. Therefore, the derive Universe model represents a model of transitioning Universe from early decelerating expansion phase to current accelerating phase, with a deceleration - acceleration redshift of $z_{t} = 0.79^{+0.04}_{-0.08}$ (OHD) and $z_{t} = 0.77^{+0.03}_{-0.04}$ (SN Ia). Fig. 4 demonstrates the variation of deceleration parameter $q$ versus redshift of the derived Universe model when it is bounding with OHD (left panel) and Pantheon compilation of SN Ia data (right panel). We obtain the present value of deceleration parameter as $q_{0} = - 0.61^{+0.075}_ {-0.045}$ (OHD) and $q_{0} = - 0.595^{+0.053}_ {-0.036}$ (SN Ia). This value of $q_{0}$ is relatively consistent with the range $q_{0} = -0.528^{+0.092}_{-0.088}$ as determine in the recent observation \cite{Gruber/2014}. Recently, Lahakare et al. \cite{Lohakare/2022} have constrained the deceleration parameter  $q_{0} = -0.69$ (OHD) and $q_{0} = -0.60$ (SN Ia) and the Universe shows a smooth transition from a decelerating phase of expansion to an accelerating phase of expansion with transition redshift of $z_{t} = 0.87$ (OHD) and $z_{t} = 0.77$ (SN Ia) respectively. It is worthwhile to note that the deceleration parameter, for instance, describes the phenomenon of the Universe which includes the phenomenon like - whether it is decelerating, accelerating, has one or more transition phases. Some important investigations that include the phenomenon of transition of the Universe from deceleration to acceleration phase are given in Refs. \cite{Farooq/2013,Capozziello/2014PRD,Yang/2020}.   
\section{Reconstruction through Energy Conditions}\label{4}
The energy conditions are the essential tools to understand the geodesics of the Universe. A complete
test of energy conditions for $f(Q)$ gravity model is explored in Ref. \cite{Mandal/2020}. Some important features of the validation of energy conditions for $f(Q)$ gravity model are discussed in De and How \cite{De/2022}. Later on Mandal et al. \cite{Mandal/2022} have showed that the extra terms to compute the energy conditions as given in Ref. \cite{De/2022}, are not going to change the energy constraint.  
In this section, we have reconstructed the symmetric teleparallel gravity on the basis of following energy conditions:
\begin{itemize}
\item Weak energy conditions (WEC) $\rho \geq 0 $, $\rho + p \geq 0$;
\item Null Energy Conditions (NEC) $\rho + p \geq 0$;
\item Dominant energy conditions (DEC) $\rho - p \geq 0$;
\end{itemize}
Collaborating with the work from Mandal et al. \cite{Mandal/2020}, the strong energy conditions (SEC) yields
\begin{equation}\label{SEC}
\rho + 3p - 6\dot{f_{Q}}H + f \geq 0.   
\end{equation}
\subsection{Reconstruction through NEC}\label{4.1}
The null energy conditions (NEC) considers future directed null vector $k^{i}$. Therefore, $T_{ij}k^{\alpha}k^{\beta} \geq 0$ leads to  
\begin{equation}\label{NEC}
    \rho+p \geq 0
\end{equation}
Eqs.(\ref{eq.15}), (\ref{eq.16}) and (\ref{NEC}) lead to 
\begin{equation}\label{NQ1}
f(Q)= \frac{c_1 Q}{H}+c_2
\end{equation}
where $c_1$ and $c_2$ are  constants.\\
Eqs. (\ref{eq.H}) and (\ref{NQ1}) lead to
\begin{equation}\label{NQ2}
f(Q)=6 c_1 H_0 \sqrt{-\Omega _{0m}+\Omega _{0m} (z+1)^3+1}+c_2
\end{equation}
Using Eq. (\ref{NQ2}) in Eq. (\ref{eq.w}), the EoS parameter $(\omega)$ is read as
\begin{equation}\label{eq.w1}
\omega= -1+\frac{6 c_1 H_0^2 \Omega _{0m} (z+1)^3}{6 c_1 H_0^2 \left(-\Omega _{0m}+\Omega _{0m} (z+1)^3+1\right)-c_2 H_0 \sqrt{\Omega _{0m} z (z (z+3)+3)+1}} 
\end{equation}
\begin{figure}[H]
\centering     
\subfigure
{\label{fig:1(a)}\includegraphics[width=70mm,]{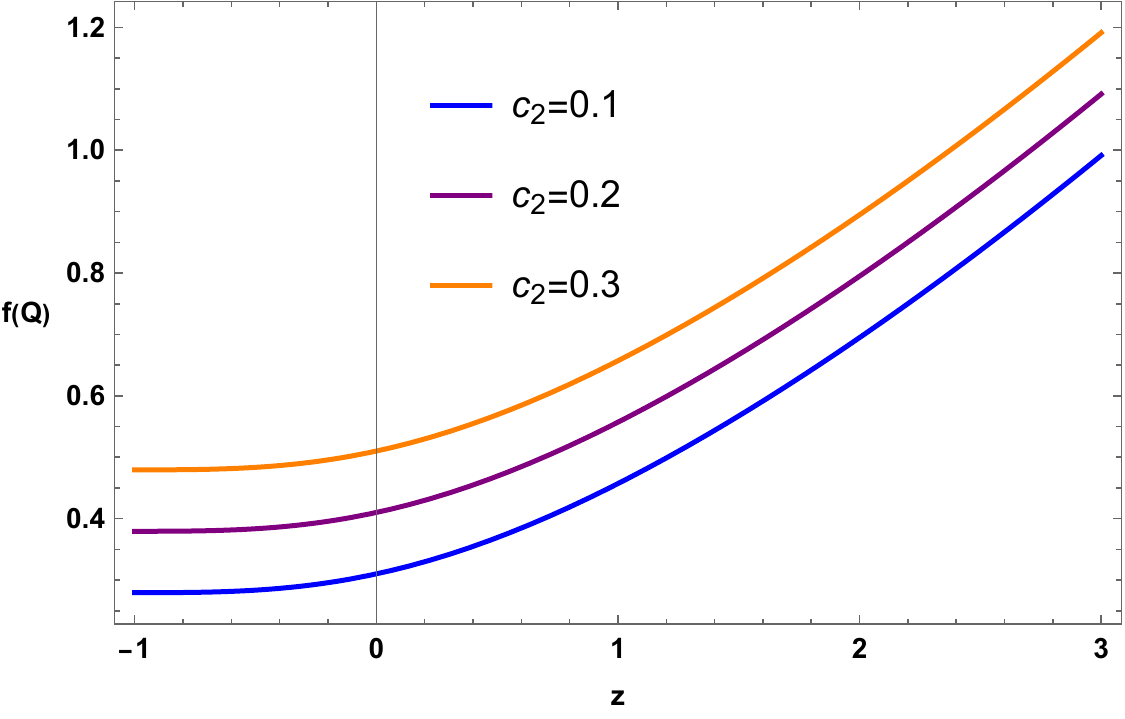}}
\hspace{1cm}
\subfigure
{\label{fig:1(b)}\includegraphics[width=70mm]{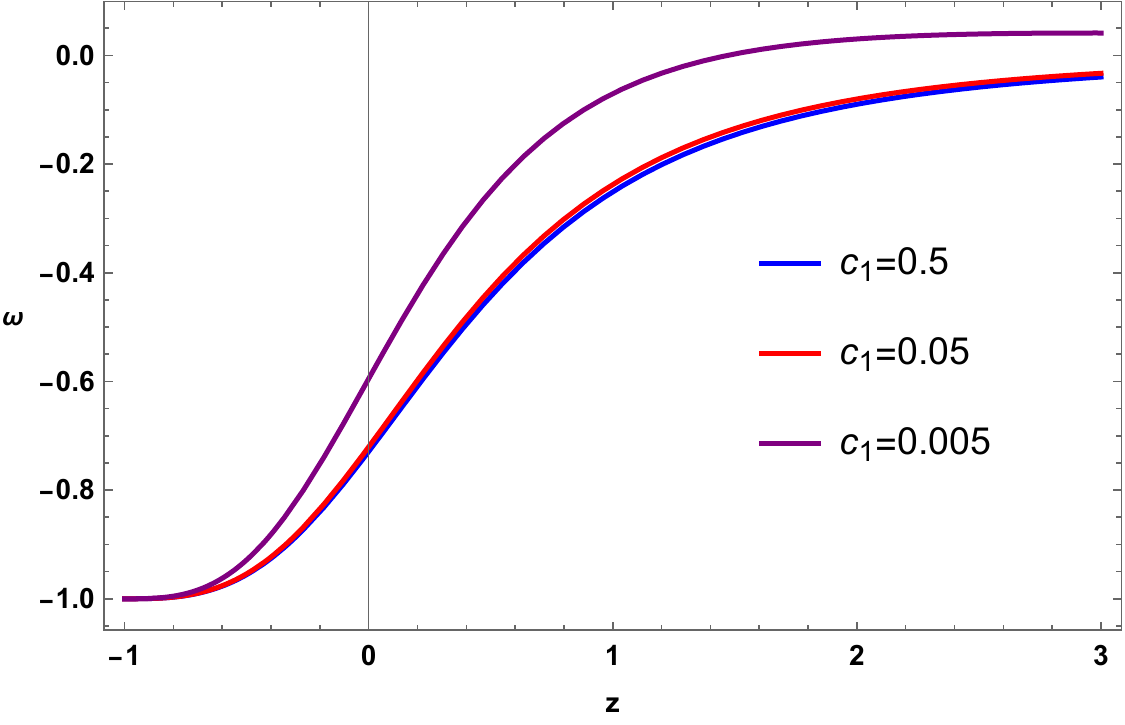}}
\caption{Evolutions of $f(Q)$ gravity versus redshift (left panel) and EoS $(\omega)$ versus redshift (right panel) for $H_0=70.02, \Omega_{0m}=0.27$.}
\label{fig-5}
\end{figure}
The evolution of $f(Q)$ versus redshift for for $H_0=70.02, \Omega_{0m}=0.27$ is shown in the left panel of Fig. \ref{fig-5} for different value of constant $c_{1}$. We observe that $f(Q)$ is decreasing function of redshift and at present it approaches a lower positive value. The evolution of EoS parameter $\omega$ versus redshift is depicted in the right panel of Fig. \ref{fig-5}. We observe that $\omega > -1$ at present time therefore the derived Universe leads the late time acceleration due to presence of quintessence energy. Furthermore, $\omega$ approaches to $-1$ at $z = -1$ $i.e.$ the derived model predicts the dominance of cosmological constant in future and it exhibits $\Lambda$CDM Universe when $z$ tends to -1.  
\subsection{Reconstruction through DEC}\label{4.2}
The dominant energy condition (DEC) ensures that the energy content of the universe is non-negative, meaning that the Universe's energy cannot be negative in any region of space. Moreover, DEC states that the matter flow along timelike or null worldlines. Therefore, an observer can measure the matter momentum density. This means that the energy dominates the other components of the energy momentum tensor being $T^{00} \geq |T^{ij}|$ and it yields the following inequality 
\begin{equation}\label{dec}
\rho - p \geq 0
\end{equation}
Eqs.(\ref{eq.15}), (\ref{eq.16}) and (\ref{dec}) yield the following expression of $f(Q)$  
\begin{equation}\label{Q1}
 f(Q)=e^{\frac{Q}{Q+\Dot{H}}-\sqrt{36 H^4+6 Q \Dot{H}+\Dot{H}^2}}c_1+e^{\frac{Q}{Q+\Dot{H}}+\sqrt{36 H^4+6 Q \Dot{H}+\Dot{H}^2}}c_2
\end{equation}
Eqs. (\ref{eq.H}) and (\ref{Q1}) lead to
\begin{eqnarray}\label{Q2}
f(Q)=c_1 e^{\frac{4 H_0^2 \left(\Omega _{0m} z \left(z^2+3 z+3\right)+1\right)}{a-b}}+c_2 e^{\frac{4 H_0^2 \left(\Omega _{0m} z \left(z^2+3 z+3\right)+1\right)}{a+b}}
\end{eqnarray}
Thus, the EoS Parameter $\omega$ is computed as 
\begin{eqnarray}
\omega=-\frac{-\frac{4 H_0^2 \Omega _0 (z+1)^3 \left(c_1 (a+b) e^{\frac{c}{a-b}}+c_2 (a-b) e^{\frac{c}{a+b}}\right)}{(a-b) (a+b) \sqrt{\Omega _0 z (z (z+3)+3)+1}}+\frac{c_1 (a-b-2 c) e^{\frac{c}{a-b}}}{a-b}+\frac{c_2 (a+b-2 c) e^{\frac{c}{a+b}}}{a+b}}{\frac{c_1 (a-b-2 c) e^{\frac{c}{a-b}}}{a-b}+\frac{c_2 (a+b-2 c) e^{\frac{c}{a+b}}}{a+b}}
\end{eqnarray}
\begin{figure}[H]
\centering     
\subfigure
{\label{fig:1(a)}\includegraphics[width=70mm,]{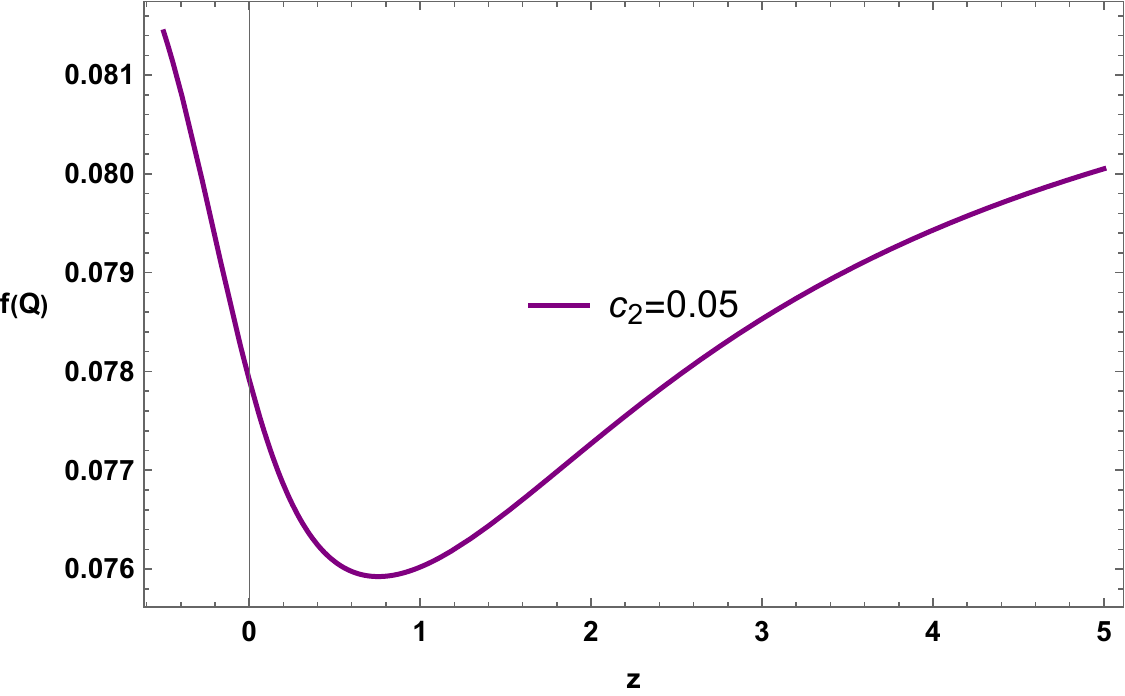}}
\hspace{1cm}
\subfigure
{\label{fig:1(b)}\includegraphics[width=70mm]{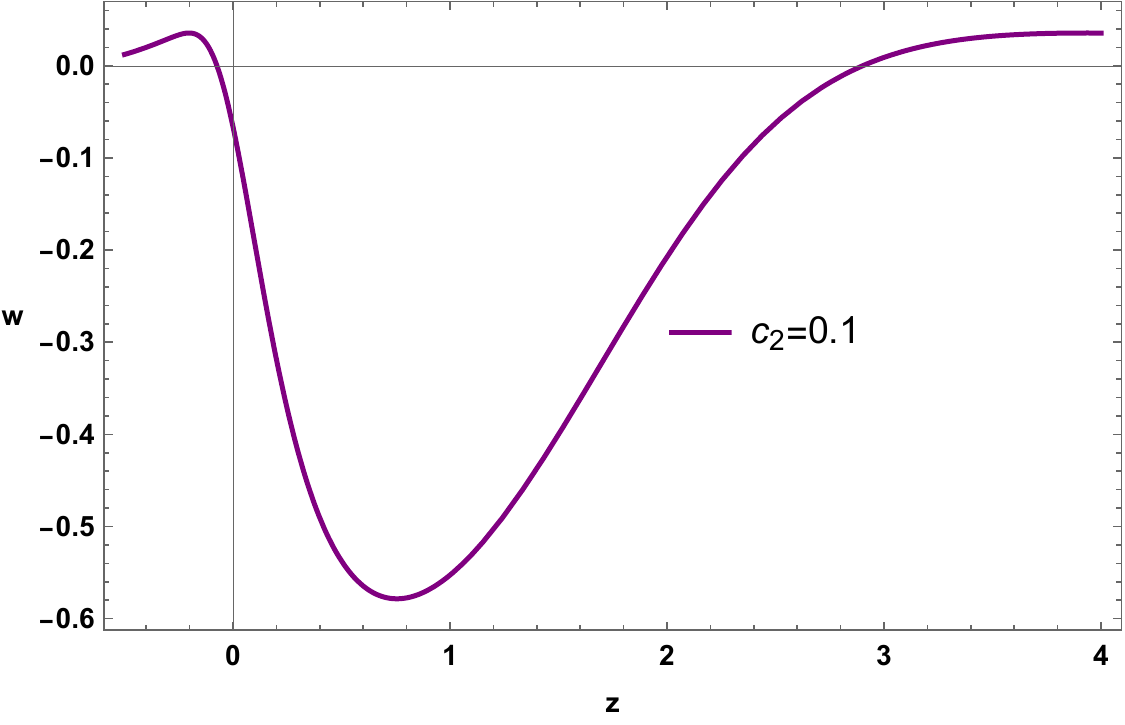}}
\caption{$f(Q)$ gravity parameter versus redshift (left panel) and EoS parameter $\omega$ versus redshift (right panel) for $H_0=70.02, \Omega_{0m}=0.27$.}
\label{fig-2}
\end{figure}
where 
\begin{eqnarray}
a&=&\frac{H_0^2 \Omega _{0m} (z+1)^3}{\sqrt{\Omega _{0m} z \left(z^2+3 z+3\right)+1}}+c;\\
b&=&\sqrt{H_0^4 \left(\frac{4 H_0^2\Omega _{0m}^2 (z+1)^6}{c}+24 \Omega _{0m} (z+1)^3 \sqrt{\frac{c}{4 H_0^2}}+16 \left(\frac{c}{4 H_0^2}\right){}^2\right)};\\
c&=&4 H_0^2 \left(\Omega _{0m} z \left(z^2+3 z+3\right)+1\right).
\end{eqnarray}
Fig. 6 depicts the graphical representation of $f(Q)$ (left panel) and EOS parameter (right panel) with respect to $z$. We observe that $f(Q)$ decreases with $z$ in past and it flip to increasing function of $z$ at present. 
\subsection{Reconstruction through SEC}\label{4.3}
 The strong energy condition places a stringent constraint on the energy content of the universe, and it is not always satisfied by standard forms of matter and energy. Violation of the strong energy condition is associated with phenomena like cosmic acceleration and the inequality relation for strong energy condition is given as
\begin{equation}
    \rho+3p \geq 0
\end{equation}
After putting the Eqs.(\ref{eq.15}) and (\ref{eq.16}) in the above equation and solving for $f(Q)$ we get 
\begin{equation}
    f(Q)=e^{\frac{Q}{Q+3\Dot{H}}-3\sqrt{36 H^4+6 Q \Dot{H}+\Dot{H}^2}} \Bigg[c_1+e^{\frac{Q\sqrt{H^4+2Q\Dot{H}+\Dot{H}^2}}{2Q\Dot{H}}}
    c_2\Bigg]
\end{equation}

\begin{eqnarray}
    f(Q)=\exp \left(\frac{2 H_0^2 \left(b^2\right)}{-\frac{a}{2}+\frac{3 H_0^2 \Omega _{0m} (z+1)^3}{2 b}+2 H_0^2 \left(b^2\right)}\right) \left(c_2 \exp \left(\frac{a b}{6 H_0^2 \Omega _{0m} (z+1)^3}\right)+c_1\right)
\end{eqnarray}

and the EoS Parameter($\omega$) is calculated as below  by putting values in Eq.(\ref{eq.w}) and the graphs for both quantities calculated are displayed below
\begin{eqnarray}
    \omega=\frac{\frac{24 H_0^2 \left(\Omega _{0m} ((2 b+1) z (z (z+3)+3)+1)+2 b\right) \left(c_1+c_2 e^d\right)}{H_0^2 \left(4 b^3+3 \Omega _{0m} (z+1)^3\right)-a b}+\frac{a c_2 e^d \left(\Omega _{0m} ((2 b+1) z (z (z+3)+3)+1)+2 b\right)}{b^2 H_0^2 \Omega _{0m} (z+1)^3}-6 \left(c_1+c_2 e^d\right)}{2 \left(3 \left(c_1+c_2 e^d\right)-H_0^2 \left(\Omega _{0m} z (z (z+3)+3)+1\right) \left(\frac{24 b \left(c_1+c_2 e^d\right)}{H_0^2 \left(4 b^3+3 \Omega _{0m} (z+1)^3\right)-a b}+\frac{a c_2 e^d}{b H_0^4 \Omega _{0m} (z+1)^3}\right)\right)}
\end{eqnarray}
\begin{eqnarray}
    a&=&\sqrt{H_0^4 \left(\frac{9 \Omega _{0m}^2 (z+1)^6}{b^2}+16 \left(b^2\right)^2+72 \Omega _{0m} (z+1)^3 b\right)}\\
    b&=&\sqrt{\Omega _{0m} z \left(z^2+3 z+3\right)+1}\\
    d&=&\frac{a b}{6 H_0^2 \Omega _{0m} (z+1)^3}
\end{eqnarray}
\begin{figure}[H]
\centering     
\subfigure
{\label{fig:1(a)}\includegraphics[width=70mm,]{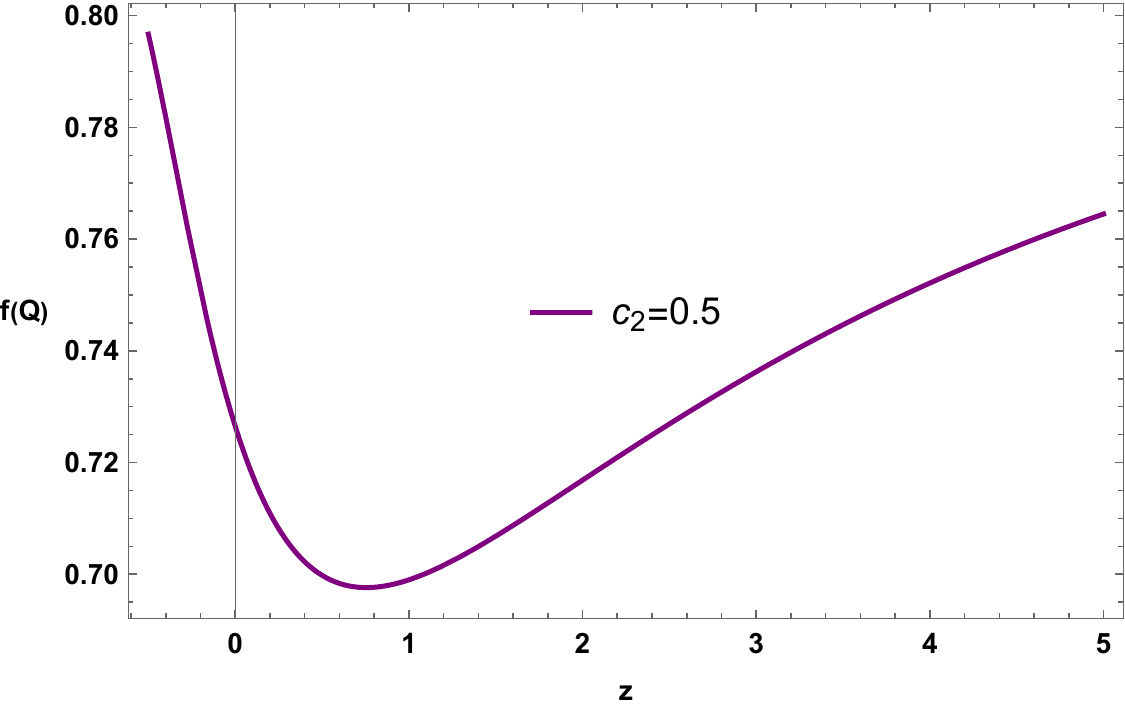}}
\hspace{1cm}
\subfigure
{\label{fig:1(b)}\includegraphics[width=70mm]{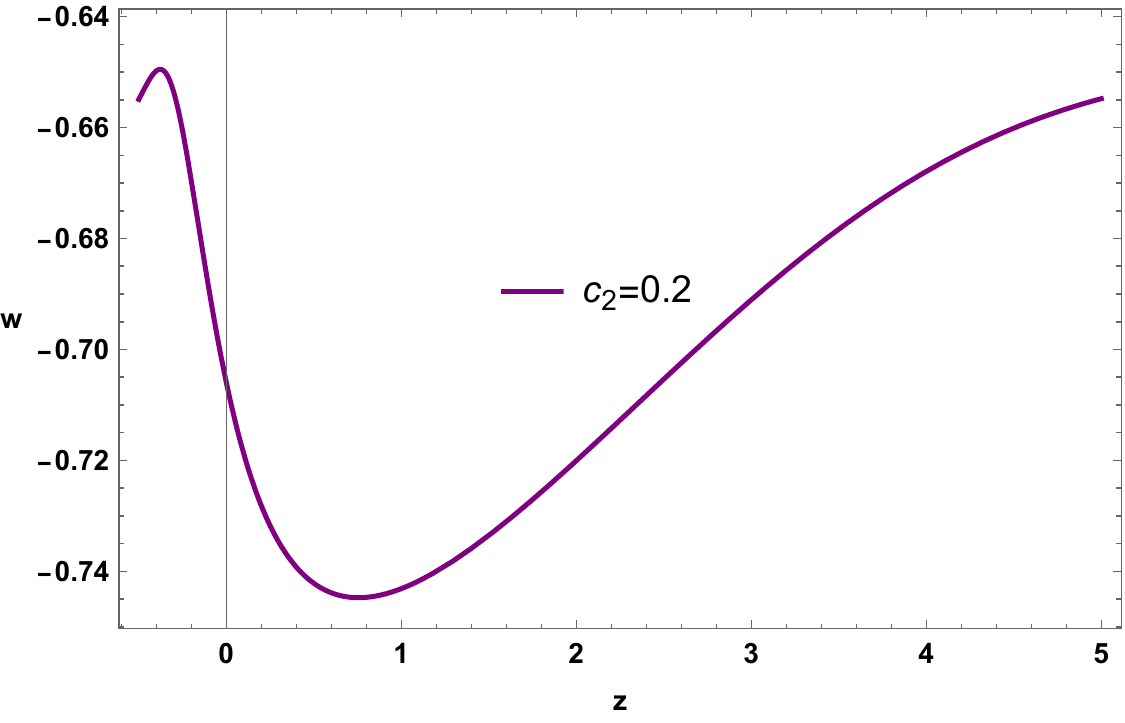}}
\caption{$f(Q)$ gravity parameter versus redshift (left panel) and EoS parameter $\omega$ versus redshift (right panel) for $H_0=70.02, \Omega_{0m}=0.27$.}
\label{fig-2}
\end{figure}
Fig. 7 shows the graphical analysis of $f(Q)$ (left panel) and EOS parameter (right panel) with respect to $z$. We observe that $f(Q)$ decreases with $z$ in past and it flip to increasing function of $z$ at present. 

\section{Conclusion}\label{5}
In this paper, we have investigated some reconstructions of $f(Q)$ gravity with energy conditions. The $f(Q)$ theory of gravity is an interesting theory that describes a fundamental approach of gravitational phenomenon. In this theory, the gravitational interaction is completely characterized by the non-metricity of $Q$. It is worthwhile to note that $f(Q)$ gravity uses the Weylian extension of Riemann's geometry, in which the metricity condition, is not valid and the non-metricity scalar $Q$ playing an analogous role to the one played by Ricci scalar in general theory of relativity. In our study, We have employed the reconstruction technique to generate explicit formulations of the $f (Q)$ Lagrangian for several types of matter sources  and computed the field equations and EoS parameter $\omega$ for the different reconstructed $f(Q)$ models on the basis of weak, null and strong energy conditions which lead the scenario of accelerating Universe, quintessence region and cosmological constant $\Lambda$. To confront the $f(Q)$ model with observations, we used 55 $H(z)$ observational data and Pantheon compilation of SN Ia data \cite{Scolnic/2018} which includes 1048 SN Ia apparent magnitude measurements in the redshift range $0.01 < z < 2.3$ to constraint the model parameters $H_{0}$ and $\Omega_{0m}$. The Universe in derived model exhibits acceleration at present epoch and it was in decelerating phase of expansion in its early time. Therefore, the model represents a transitioning Universe from early decelerating phase to current accelerating phase with transition redshifts $z_{t} = 0.87$ (OHD) and $z_{t} = 0.77$ (SN Ia) respectively. Furthermore, we estimate the current value of deceleration parameter is as $q_{0} = -0.69$ (OHD) and $q_{0} = -0.60$ (SN Ia). We also compute the present age of the universe as $t_{0} = 14.53^{+20.0}_{-0.17}$ Gyrs (OHD) and $t_{0} = 14.61^{+0.05}_{-0.10}$ Gyrs (SN Ia). Moreover, the explicit expressions of $f(Q)$ and EOS parameter $\omega$ are obtained as a function of redshift $z$ and its graphical analysis shows that $f(Q)$ decreases with $z$ in past and it flip to increasing function of $z$ at present while EOS parameter $\omega$ evolves in the range $-1 <\omega < 0$ at present epoch, therefore the derived Universe leads the late time acceleration due to presence of quintessence energy.\\

To conclude, in this paper, we have reconstructed $f(Q)$ theory of gravitation by taking into account the energy conditions and we have proven its consistency with current astrophysical observations, and as an important theoretical tool to describe the late time acceleration of the Universe without aid of dark energy. The obtained results also suggest some clue/idea for further generalizations of $f(Q)$ type models which lead an interesting geometric alternatives of dark energy.

\end{document}